\documentclass[aps,prl,reprint,superscriptaddress]{revtex4-2}

\usepackage{gensymb} % for the degree symbol
\usepackage{bbold} % for identity matrix 1
\usepackage{amsmath, bm}
\usepackage{graphicx}
\usepackage{multirow}  % for multi-row table entries
\usepackage{subfigure}  % use for side-by-side figures
\usepackage{textgreek}  % for muon symbol outside of math mode
\usepackage[sort&compress]{natbib} % for references range
\usepackage{hyperref}

\newcommand{\doHMN}[2]{%
  \begingroup\lccode`~=`#1
  \lowercase{\endgroup\let~}#2%
  \mathcode`#1="8000
}

\newcommand{\musr}{$\mu$SR}

\begin{document}

%\preprint{}

\title{Muon sites in
	PbF\textsubscript{2} and YF\textsubscript{3}: decohering environments and the role of anion Frenkel defects}


\author{J. M. Wilkinson}
\email[]{john.wilkinson@physics.ox.ac.uk}
%\thanks{}
\affiliation{Clarendon Laboratory, University of Oxford Department of Physics, Parks Road, Oxford, OX1 3PJ, United Kingdom}

\author{F. L. Pratt}
\affiliation{ISIS Facility, STFC Rutherford Appleton Laboratory, Didcot OX11 0QX, United Kingdom}

\author{T. Lancaster}
%\thanks{}
\affiliation{Centre for Materials Physics, Durham University, Durham DH1 3LE, United Kingdom}

\author{P. J. Baker}
\affiliation{ISIS Facility, STFC Rutherford Appleton Laboratory, Didcot OX11 0QX, United Kingdom}

\author{S. J. Blundell}
\email[]{stephen.blundell@physics.ox.ac.uk}
%\thanks{}
\affiliation{Clarendon Laboratory, University of Oxford Department of Physics, Parks Road, Oxford, OX1 3PJ, United Kingdom}

\date{\today}

\begin{abstract}
Muons implanted into ionic fluorides often 
lead to a so-called F--$\mu$--F state, in which the time evolution of the muon spin
contains information about the geometry and nature of the muon site. Nuclei more distant from the muon than the two nearest-neighbor fluorine ions result in
decoherence of the F--$\mu$--F system and this can yield additional quantitative information about the 
state of the muon.  We demonstrate how this can be applied to the determination of muon sites
within the ionic fluorides $\alpha$-PbF$_2$ and YF$_3$ which contain fluoride ions in different crystallographic environments.  Our results can be used to distinguish between different crystal phases and reveal the presence of anion Frenkel defects in $\alpha$-PbF$_2$.
\end{abstract}

\maketitle

%\section{}
%% Put \label in argument of \section for cross-referencing
%%\section{\label{}}
%\subsection{}
%\subsubsection{}

% What is muSR
% Limitations of muSR and DFT+mu
Muon spin rotation (\musr) is a technique which involves implanting spin-polarized
positive muons (lifetime $\tau_{\mu}=2.2$~$\mu$s) in samples to probe the local magnetic environment \cite{Blundell1999,Blundell2021}.
This technique has been applied very successfully to measure vortices in superconductors
\cite{Sonier2000}, explore the ground states in magnetic materials
\cite{Reotier1997, Pratt2011}, probe the physics of hydrogen-like defect states in semiconductors \cite{Patterson1988, Cox2009, Shimomura2016}
and in many other situations.  In order to quantitatively analyze the data obtained from \musr\
experiments, one needs to know the final stopping site of the implanted muon within 
the crystal, \emph{and} the extent to which the muon perturbs the local crystallographic 
and electronic structure. Density functional theory (DFT) calculations have recently been used to address this question \cite{Moeller2013, Bernardini2013, Bonfa2016}.   
The muon is placed at a randomly chosen site in the unit cell and the structure relaxed, with all atoms and the muon allowed to move until convergence is reached and the final energy evaluated; repeating this for many initial muon positions and identifying the minimum energy configuration yields an estimate of the muon site and allows the local distortions of the structure to be identified.  This method is often referred 
to as ``DFT+$\mu$''.

Ionic fluorides are a useful class of materials for the study of muon stopping sites and the muon-induced 
perturbation on the local crystallographic environment \cite{Moeller2013, Lancaster2007}. Fluorine, being the most 
electronegative element \cite{Allred1961}, is a very attractive atom for the incoming $\mu^+$ and, in ionic fluorides, the muon commonly stops between two fluorine anions, adopting a F--$\mu$--F state, somewhat analogous to a bifluoride ion (HF$_2^-$). Following implantation, the 
muon spin becomes entangled with the spins on nearest-neighbour fluorine nuclei, and the muon's polarization 
then evolves with time as governed by the magnetic dipolar Hamiltonian, producing a characteristic beating oscillatory signal in the measured positron asymmetry \cite{Brewer1986, Lancaster2007} that allows the muon site to be identified.
The muon's polarization also decoheres into the environment (the spin system consisting of all the other nuclei in the compound), becoming lost in an irreversible
process, which causes a relaxation in the beating oscillatory signal. It has been shown very recently that this can be quantitatively modelled to produce excellent agreement with experimental data in the simple cubic fluorides CaF$_2$ and NaF \cite{Wilkinson2020}.

In this letter, we extend the method used for CaF$_2$ and NaF \cite{Wilkinson2020}, and apply it to 
two ionic fluorides: YF\textsubscript{3} \footnote{Although YF\textsubscript{3}
has been measured and analyzed before \cite{Bernardini2013}, the crystal structure used for the 
analysis was incorrect, and it was therefore not possible to fit the data without including a
phenomenological relaxation function.} and $\alpha$-PbF\textsubscript{2}. These were chosen as they have 
structural phases with a more complicated structure than the simple cubic phases considered previously, and thereby
show that this method can be utilized in conjunction with DFT+$\mu$ to provide an insight 
into structural phase transitions and defect states, allowing us to gain a fuller understanding 
of the nature of the perturbation of the muon on the surrounding nuclei. 

% F--mu--F state theory
When a muon is implanted in a sample, it interacts with the surrounding nuclear spins by means of 
the dipole-dipole Hamiltonian $\mathcal{H}$, given by
\begin{equation}
\mathcal{H} = \sum_{i >j} \frac{\mu_0 \gamma_i \gamma_j}{4 \pi \hbar |\mathbf{r}_{ij}|^3} \Big[ \mathbf{s}_i \mathbf{\cdot} \mathbf{s}_j - 3 \big(\mathbf{s}_i \mathbf{\cdot}\mathbf{\hat{r}}_{ij} \big)\big(\mathbf{s}_j \mathbf{\cdot}\mathbf{\hat{r}}_{ij} \big) \Big],
\label{eq:MagneticDipolarHamiltonian}
\end{equation}
where $\mathbf{r}_{ij}$ is the vector linking spins $i$ and $j$, and all other symbols having their
usual meanings. For the case of a muon interacting with spin-$\frac{1}{2}$ fluorine ($^{19}$F) nuclei ($\gamma_\mathrm{F} = 2\pi \times 40.061\, \mathrm{MHz\,T}^{-1}$),
the muon`s polarization then evolves in an observable pattern of of beats (the frequencies of which 
provide information of the surrounding nuclei, due to the $\mathbf{r}_{ij}$ dependence of the 
Hamiltonian), with a relaxation which is due to the system decohering with the environment of further
nearest-neighbours, which have a weaker, but non-negligible coupling to the muon. Including 
all the nuclei in the sample directly in Eq. \eqref{eq:MagneticDipolarHamiltonian} is not possible since the dimension of $\mathcal{H}$ grows exponentially with the number of spins included.  Therefore,  following \cite{Wilkinson2020} we cut off our Hilbert space to include
enough nearest-neighbours to describe the main features of the \musr\  asymmetry, and then rescale the coupling
to the $k$ ions most distant to the muon using a parameter $\zeta_k$ which is chosen so that the second moment of our reduced system matches that of the infinite system.
The variance of the field distribution at the muon caused by $M$ spins is 
$(\sigma_M/\gamma_\mu)^2 = \frac{2}{3}(\frac{\mu_0}{4\pi})^2\hbar^2 \sum_{j=1}^M\gamma_j^2I_j(I_j+1)/r_j^6$, 
where $r_j$ is the distance from the muon to the $j$\textsuperscript{th} nucleus with spin $I_j$ and
gyromagnetic ratio $\gamma_j$, $\gamma_\mu $($=2\pi\times135.5$ MHz T$^{-1}$) is the muon gyromagnetic 
ratio, and the sum converges as $M\to\infty$. We then calculate $\zeta_k$ from
\begin{equation}
\sigma_\infty^2 = \sigma_\mathrm{nn}^2 + \frac{2}{3}\Big(\frac{\mu_0}{4\pi}\Big)^2\hbar^2\gamma_\mu^2\sum_{j \in k}\frac{\gamma_j^2I_j(I_j+1)}{(\zeta_k r_j)^6},
\label{eq:nnnbroadening}
\end{equation}
and then evaluate our exact calculation of the muon polarization to the restricted set of muon, nearest-neighbours and the set of $k$ ions (with interaction strength rescaled by $\zeta_k$).

\begin{figure*}
\centering
\includegraphics[width=0.94\textwidth]{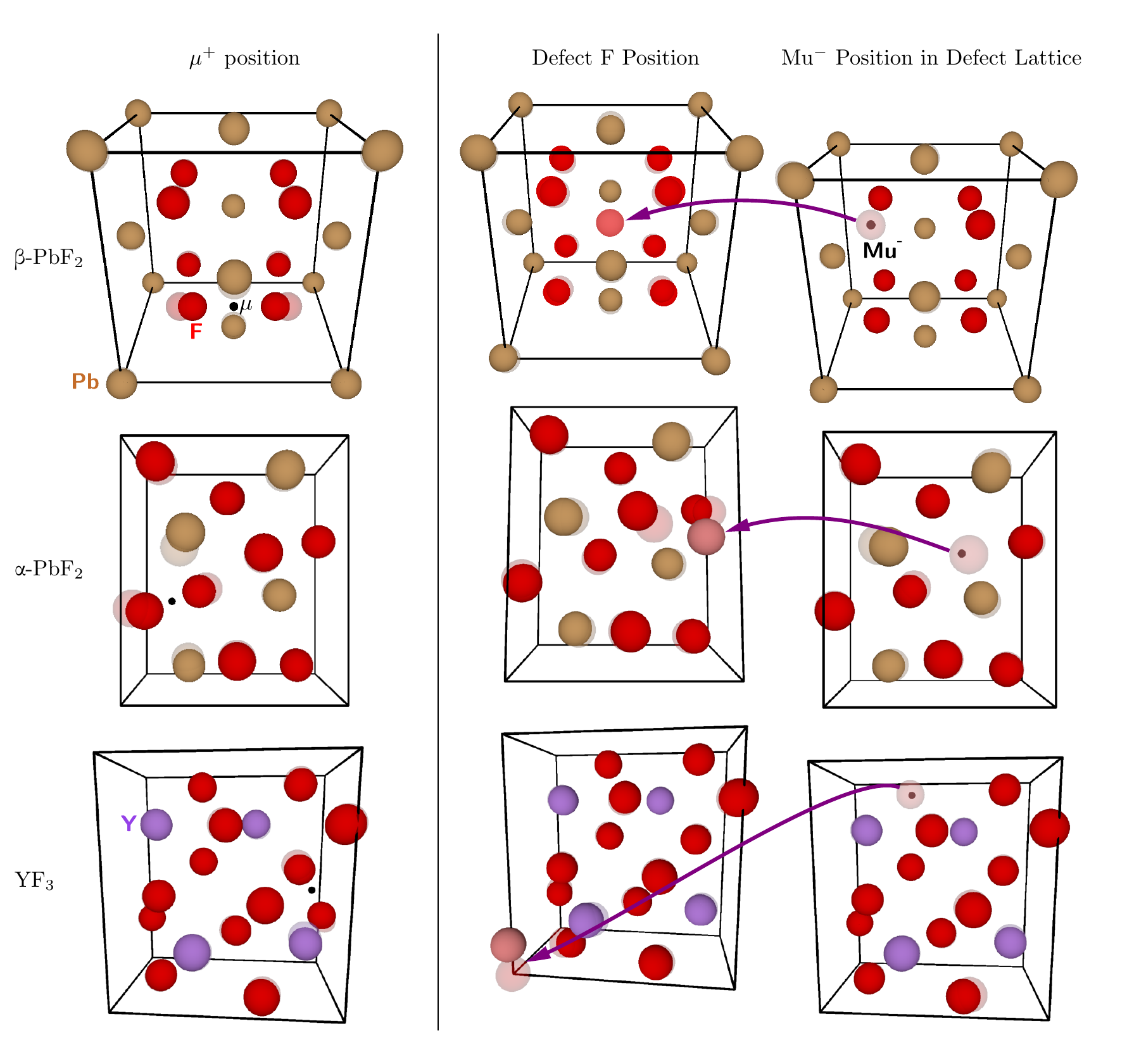}
\caption{{\bf Muon sites in PbF\textsubscript{2} and YF\textsubscript{3}}, calculated with DFT.
	The left-hand side shows the muon site calculated with the DFT+$\mu$ method, as described in the 
	text. In all cases, the muon (black sphere) sits in between two fluoride ions (red spheres), and 
	the transparent (solid) spheres show the locations of the surrounding ions before (after) the 
	perturbation of the implanted muon. The centre column shows the position of the defect fluorine 
	in the perfect lattice, and the final (initial) relaxed positions of the surrounding ions are 
	displayed as solid (transparent) spheres. Finally, the right-hand column shows the site of the Mu$^-$
	(black sphere) in the vacancy created by such a defect, and the effect of this on the surrounding ions.
}
\label{fig:muonsites}
\end{figure*}

In many fluorides, an additional relaxation component is also present in the \musr\  asymmetry. The origin 
of this component has up until now been unidentified, but we believe that in PbF\textsubscript{2} this is 
due to diamagnetic Mu$^-$ states located in anion vacancies, the origin of which are due to anion Frenkel defects 
(AFDs). The likelihood of AFDs forming is quantified by the defect formation energy, $g_\mathrm{F}$, which 
usually is of the order of a few eV, meaning that a large abundance of these defects often occur in equilibrium 
at 
temperatures of the order of hundreds of kelvin, and some can be frozen into the material and still be present at low temperatures. 
We therefore utilized DFT to estimate the anion Frenkel defect formation energy, with an approach similar 
to that undertaken before in pyrochlores \cite{Panero2004a}: We created a supercell composed of $2\times2\times2$ 
conventional unit cells, and one of the anions was displaced to a new site of high symmetry, and the cell 
relaxed. The location of the defects which had the lowest energy are shown in Fig.~\ref{fig:muonsites}. 
We found that if the anion is placed in an interstitial site sufficiently far away from the the vacancy, 
the atoms would not relax to their original positions. The final relaxed energies of the supercells containing 
the defect were compared to those without defects to obtain an estimate for $g_\mathrm{F}$, and these energies are 
tabulated in Table \ref{tab:defectenergies}, alongside the experimental values where available, which agree
well with our calculations. From this, one can conclude that both structural phases of PbF\textsubscript{2} 
are much more likely to contain Frenkel defects than YF\textsubscript{3} and  CaF\textsubscript{2}, a result
which is supported by the absence of evidence of such defect states in those compounds in our \musr\  experiments.

\begin{table}
	\begin{tabular}{ccc}
	\hline\hline
		Compound			   	&  $g_\mathrm{F}$ (eV) & $g_\mathrm{F}$ (eV) \\
		& (calculated) & (experimental) \\
		\hline
		YF\textsubscript{3}   			&  3.70                		    &   -                 \\
		CaF\textsubscript{2}   			&  2.17                		    
									&   2.2--3.1 \cite{Zhukov1998,Mackrodt1982} \\
		\textalpha-PbF\textsubscript{2} 	&  1.55                 	    
									&   1.12 \cite{Samara1979}                \\
		\textbeta-PbF\textsubscript{2}    	&  1.05		
									&   0.9--1.1 \cite{Zhukov1998,Samara1979}  \\ \hline\hline
	\end{tabular}
	\caption{Anion Frenkel defect energies $g_\mathrm{F}$ calculated by DFT (also showing the experimental values
		where available, using the method described in the text. 
		}
	\label{tab:defectenergies}
\end{table}

We obtained samples of both YF\textsubscript{3} and PbF\textsubscript{2} commercially, and used
an X-Ray diffractometer to confirm that they did not contain any significant impurities (see Supplemental 
Information) and that the PbF$_2$ sample adopted the $\alpha$-phase. 
These samples were then wrapped in silver foil measured with the MuSR spectrometer at the ISIS 
facility~\footnote{Data for this study are available from doi:10.5286/ISIS.E.RB1920547.}. The muon decay asymmetry was calculated from the number of counts in the forward and backward
detectors.

High statistics data were collected for both YF\textsubscript{3} and 
PbF\textsubscript{2} (274 and 358 million decay events, respectively).

The stopping sites of the muon were calculated using DFT+$\mu$ for YF\textsubscript{3},
and both the \textalpha\ and \textbeta\ phases of PbF\textsubscript{2} (see Supplemental 
Information for details). The sites which had the lowest overall energy 
are depicted in Fig.\ref{fig:muonsites} \footnote{There were some 
additional sites for \textalpha-PbF$_2$ and YF$_3$ which had
energy only a few meV above these, but the muon polarization
due to these sites are not realized in the data. See Supplmental
Information for details.} The stopping sites of the Mu$^-$ ions 
in the vacancies caused 
by the AFDs were also calculated and are shown in the figure; they were found to be very close to the 
positions of the anion vacancies, as expected.

% ***** PbF2 *****

\begin{figure}
	\centering
	\includegraphics[width=0.5\textwidth]{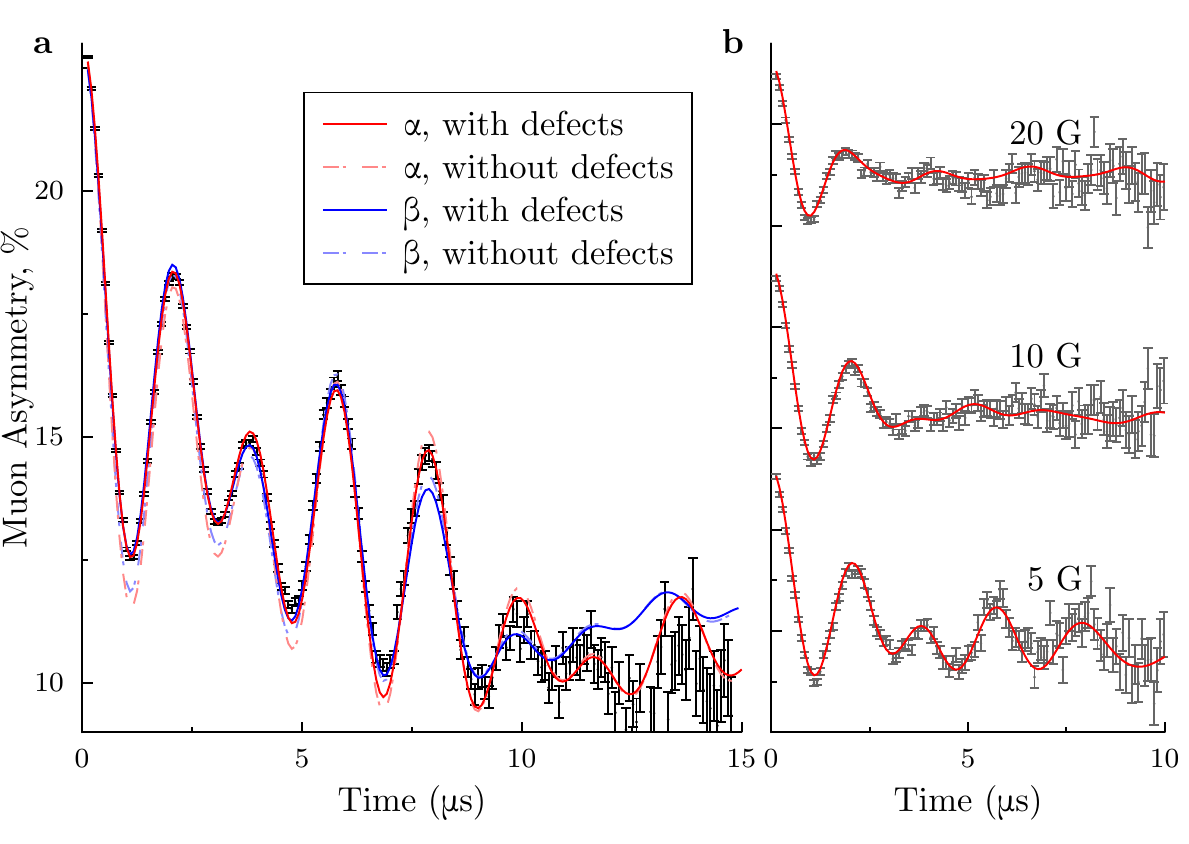}
	\caption{{\bf $\alpha$-PbF\textsubscript{2} Results.} {\bf a} shows the zero-field \musr\  data obtained 
			for our sample of PbF\textsubscript{2}, and the solid (dashed) lines corresponding
			to fits of the data with (without) taking into account the Mu$^-$ entering into a 
			defect state. {\bf b} shows the data obtained with a longitudinal field applied,
		plotted with the muon polarization as described in the text.}
	\label{fig:pbf2}
\end{figure}

The PbF\textsubscript{2} data were fitted with the function 
\begin{align}
	\label{eq:pbf2fit}
	A(t) &= A_\mathrm{r}[(1-c_\mathrm{AFD})P^\mu(r_\mathrm{nn1}, r_\mathrm{nn2}, \zeta_4; t) 
	 + c_\mathrm{AFD}P^{\mathrm{Mu}^-}(t)] \nonumber \\ & \qquad + A_\mathrm{bg}e^{-(\lambda t)^2}
\end{align}
assuming the \textalpha\ phase and also, to test the robustness of the procedure, assuming the \textbeta\ phase. The first term $P^\mu(..)$ describes the muon polarization
calculated from the dipolar Hamiltonian \eqref{eq:MagneticDipolarHamiltonian}, using the muon site calculated 
with DFT+$\mu$, where $r_\mathrm{nn1}$ and $r_\mathrm{nn2}$ represent the distance from the muon to the two 
nearest neighbour fluorines. The second term $P^{\mathrm{Mu}^-}(t)$ is the polarization of the negatively 
charged Muonium ion in an AFD site, and the final term $e^{-(\lambda t)^2}$ represents a slow relaxation
of the muon polarization due to the weak nuclear moments in the Ag sample holder. 
As Pb has a very weak moment (the only isotope with spin, $^{207}$Pb,
has $\mu=0.584\mu_\mathrm{N}$ with 23\% abundance), only the nearest
ten fluorine nuclei were included in the calculation 
of the muon polarisation (see the Supplemental Information for the positions of the included nuclei)
and so the Hamiltonian was described by a matrix of size
$2048 \times 2048$. 

The results of the fits to both the \textalpha\ and \textbeta\ phases are shown in Fig.~\ref{fig:pbf2}(a), along
with simulations for which $c_\mathrm{AFD}$ was fixed to zero (i.e.\ ignoring AFDs). From this, one can see that the best fit is
obtained including the presence of AFDs and with PbF$_2$ adopting the \textalpha\ phase.  The superiority of the \textalpha-phase fit is especially apparent for the data at longer times ($>8$ \textmu s), where the polarization
function strongly deviates from the data for the \textbeta\ phase. Further evidence supporting the validity of the model applying to the 
\textalpha\ phase of the compound is obtained by considering the second moment rescaling factor $\zeta$. The fit for the 
\textalpha\ phase obtained a value of $\zeta_4 = 0.834(5)$, very close to the calculated value of $0.8433$. 
However, for the \textbeta\ phase, the fitted value of $\zeta_\mathrm{nnn} = 0.804(4)$, strongly deviating from
the calculated value of $0.8956$. For the \textalpha\ phase, the final obtained values of $r_\mathrm{nn1}$ and 
$r_\mathrm{nn2}$ were 1.1419(4)~\AA\,and 1.2601(7)~\AA, close to the corresponding values of 1.10 and 1.21 \AA\, 
obtained with DFT+$\mu$.  These results therefore demonstrate that the details of the oscillatory signal are able to correctly distinguish between the two related crystallographic phases of PbF$_2$, but note that the sensitivity of \musr\ data to these local differences is only revealed clearly in the late-time data (obtained well after $\approx 5\tau_{\mu}$ where, because of the muon decay, the data rate is more than two orders of magnitude lower than that obtained immediately after implantation).  This result highlights the necessity for obtaining high-statistics \musr\ data so that 
these small late-time features in the data can be carefully resolved.

The best fit to the zero-field data was obtained assuming that
$9.2(2) \%$ of the diamagnetic fraction of the signal in $\alpha$-PbF$_2$ was due to AFD states. In order to test 
for the presence of these states, we applied a series of small longitudinal fields to the sample. Applying such 
fields adds an additional Zeeman term to the Hamiltonian, which, for AFD Mu$^-$ states, tends to be much larger
than the dipole interactions, and hence $P^{\mathrm{Mu}^-}(t)\approx 1$. These results, along with plots of the 
polarization function for \textalpha-PbF\textsubscript{2}, using parameters obtained from the aforementioned
fit and with no further adjustment of parameters, apart from including the longitudinal field, are shown in Figure \ref{fig:pbf2}(b). One can see that our model continues to describe the data very well, 
and the field indeed removes much of the effect of the defect states on the polarization.

% ***** YF3 *****
\begin{figure}
	\centering
	\includegraphics[width=0.5\textwidth]{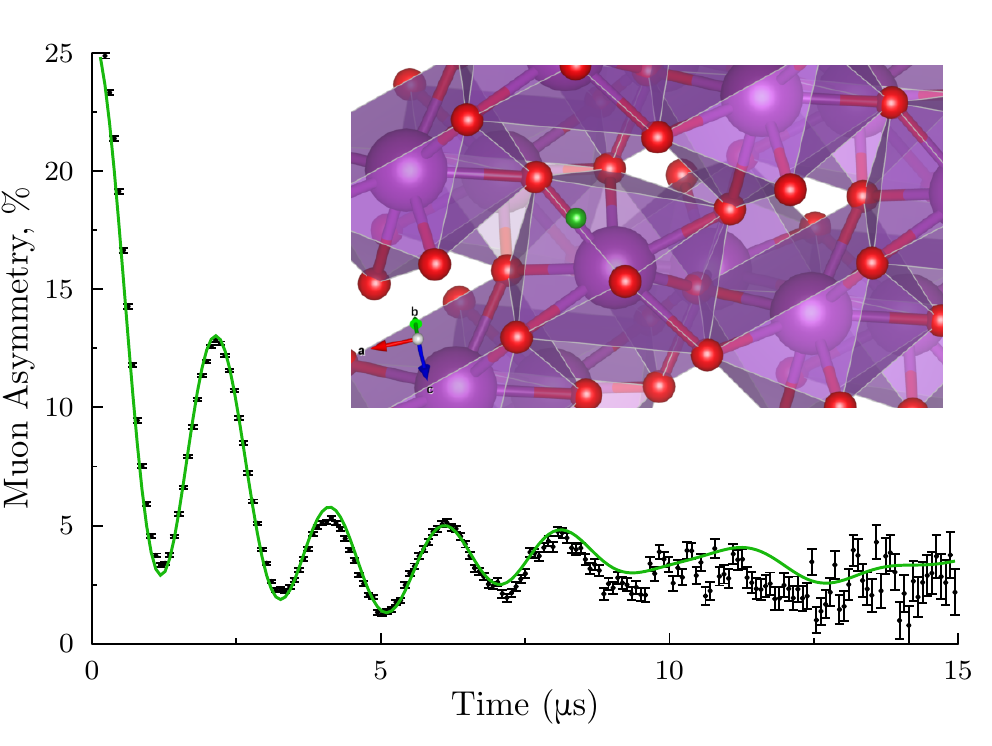}
	\caption{{\bf YF\textsubscript{3} Results}: The results of our \musr\  experiment, with the fitted 
		muon polarization using the muon site obtained by DFT. This
		site is displayed on the YF\textsubscript{3} crystal structure 
		in the inset, with the atom positions unperturbed for simplicity.}
	\label{fig:yf3}
\end{figure}

As our data for YF\textsubscript{3} did not show any slow relaxing background, and the DFT results show that AFDs
are less likely to form than in PbF\textsubscript{2}, we did not need to include AFDs in our fits of the data. Therefore we fit our YF\textsubscript{3} data to a simpler functional form, given by
\begin{equation}
	\label{eq:yf3it}
	A(t) = A_\mathrm{r}P^\mu(r_\mathrm{nn1}, r_\mathrm{nn2}, r_\mathrm{nnn1}, r_\mathrm{nnn2}, \zeta_6; t) 
		 + A_\mathrm{bg},
\end{equation}
where $r_\mathrm{nnn1}$ and $r_\mathrm{nnn2}$ are the distances 
between the muon and the two next-nearest neighbour 
fluorine nuclei, and all other symbols have the same meaning as before. 
As with PbF$_2$, the ten nearest fluorine nuclei were included in the
calculation (Y has only one natural isotope, $^{89}$Y, with a negligble
moment of $0.137\mu_\mathrm{N}$), leading again to matrices of size
$2048\times2048$ (see the Supplemenetal Information for the details of the
included nuclei).
The site which was predicted by DFT to have the lowest energy (using both the LDA and PBE functionals) 
was used to calculate the muon polarization, the fit of which is depicted as the green line in Figure 
\ref{fig:yf3}. The values obtained from the fit, and their DFT counterparts,
are tabulated in Table \ref{tab:yf3results}, where it can be seen 
that the nearest- and next-nearest- neighbour perturbations of the 
fluorines due to the muon are in excellent agreement. Note that
the discrepancy between the DFT values and the fitted values increases
with the muon-fluorine distance, and that DFT slightly underestimates the bond
lengths for nuclei close to the muon, and slightly overestimates these for those
nuclei further away. 
While the PBE functionals tend to slightly underestimate bond lengths,
the effect further from the muon may be due to the size of the supercell.

\begin{table}
	\centering
	\begin{tabular}{c  c  c  c}
	\hline\hline
		Parameter	& Experimental 		& DFT$+\mu$ 	& Difference \\
		\hline
		$r_\mathrm{nn1}$ (\AA)      & 1.173(1)    & 1.10       & $+0.073$   \\
		$r_\mathrm{nn2}$ (\AA)      & 1.278(2)    & 1.22       & $+0.058$   \\
		$r_\mathrm{nnn1}$ (\AA)     & 2.24(3)     & 2.15       & $+0.09$    \\
		$r_\mathrm{nnn2}$ (\AA)     & 2.40(1)    & 2.49      & $-0.09$    \\ 
		$\zeta_6$              & 0.849(5)      & 0.907         & $-0.058$
		\\
		\hline\hline
	\end{tabular}
	\caption{Muon-induced structural distortions in YF\textsubscript{3},
		comparing the values obtained with the fit to experimental data with the values obtained using DFT$+\mu$.}
	\label{tab:yf3results}
\end{table}

% Conclusion

In conclusion, we have shown how the analysis of \musr\  data on complex fluorides allows one to obtain a 
wealth of useful information about the muon stopping site and the environment of the muon, information
which was previously not obtainable when phenomenological relaxation functions were used to analyze
the data. We have also shown how the longer-time data are particularly useful for understanding the 
nature of the muon site. Additionally, we have shown that it is possible
to measure anion Frenkel defects using $\mu$SR in a model system, an
important result which can be extended to the analysis of magnetic
systems, where the effects of defects on the electronic structure could be
important.

We acknowledge support from EPSRC (Grant No.\ EP/N023803/1) and the use of the University of Oxford Advanced Research Computing (ARC) facility in carrying out this work
(\url{http://dx.doi.org/10.5281/zenodo.22558}). 
We are grateful to D.~Prabhakaran for performing the x-ray measurement and to Marina Filip for useful discussions. Part of this work was carried out at the STFC-ISIS muon facility.

\bibliography{references}

\end{document}